\documentclass[dvips]{article}
\usepackage{icrctc07}

\newcommand{\arcmin}{\hbox{$^\prime$}}

\newcommand{\gr}{$\gamma$-ray}

\newcommand{\degr}{\hbox{$^\circ$}}
\newcommand{\hess}{H.E.S.S.}

\newcommand{\rxj}{RX~J1713.7$-$3946}

\title{Primary particle acceleration above 100 TeV in the shell-type
Supernova Remnant RX J1713.7--3946 with deep H.E.S.S. observations}
\shorttitle{Particle acceleration above 100 TeV in the SNR RX
J1713.7--3946 with H.E.S.S.}

\authors{
  D.~Berge$^{1,2}$,
  F.~Aharonian$^{2,3}$,
  W.~Hofmann$^{2}$,
  M.~Lemoine-Goumard$^{4}$,
  O.~Reimer$^{5}$,
  G.~Rowell$^{6}$,
  H.J.~V\"olk$^{2}$,
  for the H.E.S.S.\ Collaboration
}

\shortauthors{Berge et al.}

\afiliations{
  $^1$ CERN PH Department, CH-1211 Geneva 23, Switzerland\\
  $^2$Max-Planck-Institut f\"ur Kernphysik, P.O. Box 103980, D-69029
  Heidelberg, Germany\\
  $^3$Dublin Institute for Advanced Studies, 5 Merrion Square,
  Dublin 2, Ireland\\
  $^4$Laboratoire Leprince-Ringuet, IN2P3/CNRS, Ecole Polytechnique,
  F-91128 Palaiseau,\\France\\
  $^5$Stanford University, HEPL $\&$ KIPAC, CA 94305-4085, USA\\
  $^6$School of Chemistry \& Physics, University of Adelaide, Adelaide
  5005, Australia
}
\email{berge@cern.ch}

\abstract{The shell-type supernova remnant RX J1713.7--3946 was
observed during three years with the H.E.S.S. Cherenkov telescope
system. The first observation campaign in 2003 yielded the first-ever
resolved TeV gamma-ray image. Follow-up observations in 2004 and 2005
revealed the very-high-energy gamma-ray morphology with unprecedented
precision and enabled spatially resolved spectral studies. Combining
the data of three years, we obtain significantly increased statistics
and energy coverage of the gamma-ray spectrum as compared to earlier
H.E.S.S. results. We present the analysis of the data of different
years separately for comparison and demonstrate that the telescope
system operates stably over the course of three years. When combining
the data sets, a gamma-ray image is obtained with a superb angular
resolution of 0.06 degrees. The combined spectrum extends over three
orders of magnitude, with significant gamma-ray emission approaching
100 TeV. For realistic scenarios of very-high-energy gamma-ray
production, the measured gamma-ray energies imply efficient particle
acceleration of primary particles, electrons or protons, to energies
exceeding 100 TeV in the shell of RX J1713.7--3946.}

\begin{document}
\maketitle
%Begin the section.

\section{Introduction}
\vspace{-0.4cm}
%%%%%%%%%%%%%%%%%%%%%%%%%%%  Fig3   %%%%%%%%%%%%%%%%%%%%%%%%%%%%%%%%%

\begin{figure*}
  \begin{center}
    \includegraphics [width=0.78\textwidth]{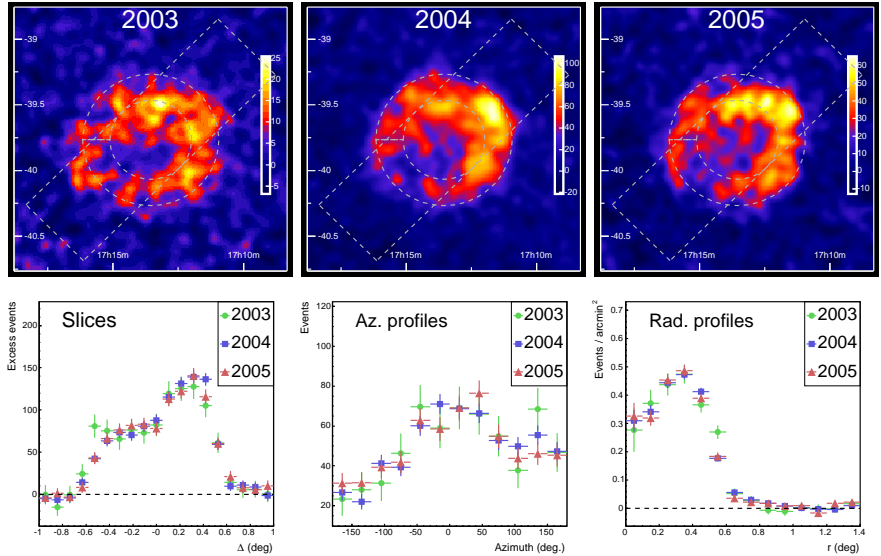}
  \end{center}
  \caption{\underline{\textbf{Upper panel:}} H.E.S.S.\ gamma-ray
    excess images from the region around RX~J1713.7$-$3946\ are shown
    for three years. \underline{\textbf{Lower panel:}} 1D
    distributions generated from the non-smoothed,
    acceptance-corrected gamma-ray excess images.  }
  \label{fig1}
\end{figure*}

 %    The images are corrected for the decline of the
 %    system acceptance with increasing distance to the SNR centre and
 %    smoothed with a Gaussian of $2\arcmin$. The linear colour scale
 %    is in excess counts per smoothing radius. The dashed box
 %    (dimensions $2\degr \times 0.6\degr$) and ring ($r_1 = 0.3\degr$,
 %    $r_2 = 0.5\degr$) are used for obtaining the 1D distributions
 %    shown in the lower panel. 
 %

%     \textbf{Left:} Slices taken within the rotated dashed box
%     running through the SNR region. \textbf{Middle:} Azimuth profiles
%     integrated in a thick ring covering the shell of
%     RX~J1713.7$-$3946. \textbf{Right:} Radial profiles around the
%     centre of the SNR.

%$0\degr$ corresponds to the west part of the shell, $90\degr$ is
%    north or upward, $-90\degr$ is south or downward. 

%%%%%%%%%%%%%%%%%%%%%%%%%%%%%%%%%%%%%%%%%%%%%%%%%%%%%%%%%%%%%%%%%%%%%%%%%

The energy spectrum of cosmic rays measured at Earth exhibits a
power-law dependence over a broad energy range. Starting at a few GeV
$(1~\mathrm{GeV} = 10^9~\mathrm{eV})$ it continues to energies of at
least $10^{20}~\mathrm{eV}$. The power-law index of the spectrum
changes at two characteristics energies: in the region around $3
\times 10^{15}~\mathrm{eV}$ -- the \emph{knee} region -- the spectrum
steepens, and at energies beyond $10^{18}~\mathrm{eV}$ it hardens
again. Up to the knee, cosmic rays are believed to be of Galactic
origin, accelerated in shell-type supernova remnants (SNRs) --
expanding shock waves initiated by supernova
explosions~\cite{HillasReview}. However, the experimental confirmation
of an SNR origin of Galactic cosmic rays is difficult due to the
propagation effects of charged particles in the interstellar
medium. The most promising way of proving the existence of high-energy
particles in SNR shells is the detection of very-high-energy (VHE)
gamma rays ($E > 100~\mathrm{GeV}$), produced in interactions of
cosmic rays close to their acceleration site~\cite{DAV}.

Recently the VHE gamma-ray instrument H.E.S.S. has detected two
shell-type SNRs, RX~J1713.7$-$3946~\cite{Hess1713a,Hess1713b} and
RX~J0852.0--4622~\cite{HessVelaJr_a,HessVelaJr_b}. The two objects
show an extended morphology and exhibit a shell structure, as expected
from the notion of particle acceleration in the expanding shock
fronts.
% Both reveal gamma-ray spectra that can be described by a hard power
% law (with photon index $\Gamma \sim 2.0$) over a broad energy
% range. For RX~J1713.7$-$3946\ significant deviations from a pure power
% law at larger energies are measured~\cite{Hess1713b}.
While it is difficult to attribute the measured VHE gamma rays
unambiguously to nucleonic cosmic rays (rather than to cosmic
electrons), the measured spectral shapes favour indeed in both cases a
nucleonic cosmic-ray origin of the gamma
rays~\cite{Hess1713b,HessVelaJr_b}.

%   In the case of RX~J1713.7$-$3946\ in addition a narrow shock
%   filament seen in X-rays~\cite{HiragaXMM} indicates strong
%   amplification of the magnetic field at least in one region of the
%   rim~\cite{BerezhkoVoelk}. If such an amplified magnetic field exists
%   throughout the main volume of the SNR~--~the region for which VHE
%   gamma-ray data is presented here~--~and if consequently high
%   magnetic field values are found not only in one shock filament, but
%   on a large part of the shock surface, a leptonic origin of the VHE
%   gamma rays becomes increasingly unlikely just based on the absolute
%   level of X-ray and gamma-ray flux of
%   RX~J1713.7$-$3946~\cite{Hess1713b}.
% 
Apart from the first unambiguous proof of multi-TeV particle
acceleration in SNRs, the question of the highest observed energies
remains an important one. Only the detection of gamma rays with
energies of 100~TeV provides experimental proof of acceleration of
primary particles, protons or electrons, to the \emph{knee} region
(1~PeV). Here we present a combined analysis of H.E.S.S.\ data of
RX~J1713.7$-$3946\ of three years, from 2003 to 2005. A comparison of
the three data sets demonstrates the expected steady emission of the
source as well as the stability of the system. Special emphasis is
then devoted to the high-energy end of the combined spectrum.
%\footnote{The analysis
%presented here is already published, where appropriate we condensed
%the information and would like to refer the reader for more details to
%the original publication~\cite{Hess1713c}.}

\section{\hess\ observations}
\vspace{-0.4cm}
The High Energy Stereoscopic System (H.E.S.S.) consists of four
identical Cherenkov telescopes that are operated in the Khomas
Highland of Namibia. Its large field of view of $\approx 5\degr$ make
H.E.S.S.\ the currently best suited experiment in the field for the
study of extended VHE gamma-ray sources such as young Galactic SNRs.

The H.E.S.S.\ observation campaign of RX~J1713.7$-$3946\ started in
2003. The data were recorded during the commissioning phase of the
telescope system, with 2 out of the 4 telescopes operational. The data
set revealed extended gamma-ray emission resembling a shell
structure. It was actually the first ever resolved image of an
astronomical source obtained with VHE gamma rays. In 2004,
observations were conducted with the full telescope array. The
H.E.S.S.\ data enabled analysis of the gamma-ray morphology and the
spectrum of the remnant with unprecedented precision. A very good
correlation was found between the X-ray and the gamma-ray image. The
differential spectrum showed deviations from a pure power law at high
energies. The 2005 observation campaign was aiming at extending the
energy coverage of the spectrum to as high energies as
possible. Therefore the observations were preferentially pursued at
zenith angles larger than in the two years before to make use of the
drastically increased effective collection area of the experiment at
high energies. The analysis of these data are presented in the
following (a more detailed discussion of this analysis can be found in
\cite{Hess1713c}).

\section{Analysis results}
\vspace{-0.4cm}

%%%%%%%%%%%%%%%%%%%%%%%%%%%  Fig2   %%%%%%%%%%%%%%%%%%%%%%%%%%%%%%

\begin{figure}
  \begin{center}
    \noindent
    \includegraphics [width=0.38\textwidth]{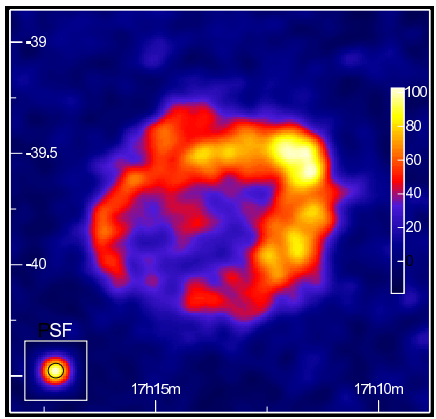}
  \end{center}
  \caption{Combined H.E.S.S.\ image of the SNR \rxj\ from the 2004 and
      2005 data. A simulated point source (\emph{PSF}) is also shown.}
  \label{fig2}
\end{figure}
%The acceptance-corrected gamma-ray excess image.

%  Note that for the 2005 data, only data recorded at zenith angles
%  less than $60\degr$ are taken into account. 

% The image is smoothed as in Fig.~\ref{fig1},

%%%%%%%%%%%%%%%%%%%%%%%%%%%%%%%%%%%%%%%%%%%%%%%%%%%%%%%%%%%%%%%%%%%%%%%%%

%%%%%%%%%%%%%%%%%%%%%%%%%%%  Fig3   %%%%%%%%%%%%%%%%%%%%%%%%%%%%%%%%%

\begin{figure*}
  \begin{center}
    \includegraphics [width=0.8\textwidth]{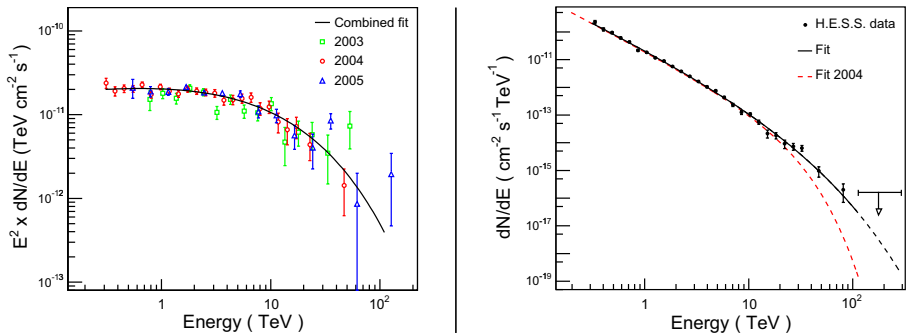}
  \end{center}
  \caption{\textbf{Left:} Comparison of H.E.S.S.\ energy-flux spectra
    of three years. The black curve is the best fit of a power law
    with exponential cutoff to the combined data, as shown on the
    \textbf{right}, where the combined \hess\ \gr\ spectrum of \rxj\
    is shown. The data are well described by the fit function, which
    is continued as dashed line beyond the fit range for
    illustration. The arrow is a model-independent upper limit,
    determined in the energy range from 113 to 300~TeV.}
  \label{fig3}
\end{figure*}

% It is the best fit of a power law with exponential cutoff to the
% combined data, where the cutoff is taken to the power of $\beta =
% 0.5$: $dN / dE = I_0 \, E^{-\Gamma} \,
% \exp\left(-(E/E_\mathrm{c})^{~\beta=0.5}\right)$.

%  The energy threshold of $\sim1$~TeV in the 2003 data is due to the
%  two-telescope operation mode and the application of a stringent cut
%  on the minimum camera image size.

%  the degradation of the optical efficiency of the system. 

% quote fit result in the text!!!

% The spectra are shown in an energy-flux representation -- flux
% points have been multiplied by $E^2$. 

%%%%%%%%%%%%%%%%%%%%%%%%%%%%%%%%%%%%%%%%%%%%%%%%%%%%%%%%%%%%%%%%%%%%%%%%%

The analysis techniques used here are presented in detail
elsewhere~(\cite{Hess2155,BackgroundPaper}). The gamma-ray
morphology measured in three years is seen in the upper panel of
Fig.~\ref{fig1}. The images are readily comparable. Very similar
angular resolutions are achieved for all years. Good agreement is
achieved, as can also be seen from the 1D distributions shown in the
lower panel, where also the statistical errors are plotted. Shown from
left to right are a slice along a thick box (cf.\ Fig.~\ref{fig1},
upper panel), an azimuthal profile of the shell region, and a radial
profile. All the distributions are generated from the non-smoothed,
acceptance-corrected excess images. Clearly, there is no sign of
disagreement or variability, the H.E.S.S.\ data of three years are
well compatible with each other.
  
The combined H.E.S.S.\ image is shown in Fig.~\ref{fig2}. Data of 2004
and 2005 are used for this Gaussian smoothed~($\sigma=2\arcmin$),
acceptance-corrected gamma-ray excess image. In order to obtain
optimum angular resolution, a special high-resolution analysis is
applied here. Besides choosing only well reconstructed events, the cut
on the minimum event multiplicity is raised to three telescopes,
disregarding the 2003 data. Moreover, an advanced reconstruction
method is chosen, \emph{algorithm~3} of \cite{HofmannShowerReco}. The
image corresponds to 62.7 hours of observation time. 6702 gamma-ray
excess events are measured with a statistical significance of
$48\sigma$. An angular resolution of $0.06\degr$ ($3.6\arcmin$) is
achieved. The image confirms nicely the published H.E.S.S.\
measurements~\cite{Hess1713a,Hess1713b}, with 20\% better angular
resolution and increased statistics. The shell of RX~J1713.7$-$3946,
somewhat thick and asymmetric, is clearly visible and almost
closed. The gamma-ray brightest parts are located in the north and
west of the SNR.

The gamma-ray spectra measured with H.E.S.S.\ in three consecutive
years are compared to each other in Fig.~\ref{fig3}~(left). In order
to compare the data, a correction for the variation of optical
efficiency of the telescopes over the years must be
applied~\cite{HessCrab}. After that correction, very good agreement is
found. The measured spectral shape remains unchanged over time. The
absolute flux levels are well within the systematic uncertainty of
20\%. As expected for an object like RX~J1713.7$-$3946, no flux
variation is seen on yearly timescales. Clearly, the performance of
the telescope system is under good control.

The combined data of three years are shown in
Fig.~\ref{fig3}~(right). This energy spectrum of the whole SNR region
corresponds to $91$~hours of H.E.S.S.\ observations. The combined
spectrum extends over almost three decades in energy beyond 30~TeV,
and is compatible with previous H.E.S.S.\ measurements. Taking all
events with energies above 30~TeV, the cumulative significance is
$4.8~\sigma$. Different spectral models can be fit to the data. A pure
power law is clearly ruled out, alternative spectral models like a
power law with exponential cutoff, a broken power law, or a power law
with energy-dependent index, all provide significantly better
descriptions of the data, but none of these alternative models is
favoured over the other.

\section{Summary}
\vspace{-0.4cm}
The complete H.E.S.S.\ data set of the SNR RX~J1713.7$-$3946\ recorded
from 2003 to 2005 is presented here. Very good agreement is found for
both the gamma-ray morphology and the differential energy spectra over
the years. The combined analysis confirms the earlier findings nicely:
the gamma-ray image reveals a thick, almost circular shell with
significant brightness variations. The spectrum follows a hard power
law with significant deviations at higher energies (beyond a few TeV).

In the combined image using $\sim63$~hours of H.E.S.S.\ observations
an unprecedented angular resolution of $0.06\degr$ is achieved. The
high-energy end of the combined spectrum approaches 100~TeV with
significant emission $(4.8\sigma)$ beyond 30~TeV. Given the systematic
uncertainties in the spectral determination at these highest energies
and comparable statistical uncertainties despite the long exposure
time, this measurement is presumably close to what can be studied with
the current generation of imaging atmospheric Cherenkov telescopes.

From the largest measured gamma-ray energies one can estimate the
corresponding energy of the primary particles. In case of
$\pi^0$-decay gamma rays, energies of 30~TeV imply that primary
protons are accelerated to $30~\mathrm{TeV} / 0.15 = 200~\mathrm{TeV}$
in the shell of RX~J1713.7$-$3946. On the other hand, if the gamma
rays are due to Inverse Compton scattering of VHE electrons, the
electron energies at the current epoch can be estimated in the
Thompson regime as
$E_\mathrm{e}~\approx~20~\sqrt{E_\gamma}~\mathrm{TeV} \approx
110~\mathrm{TeV}$. At these large energies Klein--Nishina effects
start to be important and reduce the maximum energy slightly such that
$\sim100~\mathrm{TeV}$ is a realistic estimate.

RX~J1713.7$-$3946\ remains an exceptional SNR in respect of its VHE
gamma-ray observability, being at present the remnant with the widest
possible coverage along the electromagnetic spectrum. The H.E.S.S.\
measurement of significant gamma-ray emission beyond 30~TeV without
indication of a termination of the high-energy spectrum provides proof
of particle acceleration in the shell of RX~J1713.7$-$3946\ beyond
$10^{14}$~eV, up to energies which start to approach the region of the
cosmic-ray \emph{knee}.

\bibliography{icrc0524}
\bibliographystyle{plain}
\end{document}